\documentclass{revtex4}
\usepackage{graphicx}

\begin{document}

\title{Pressure induced Insulator-Metal transition in LaMnO$_3$}

\author{Javier D. Fuhr and Blas Alascio} 
\address{Instituto Balseiro and Centro At\'omico Bariloche, CNEA, 8400
  San Carlos de Bariloche, Argentina.}
\author{Michel Avignon} 
\address{Institut N\'eel, CNRS and Universit\'e Joseph Fourier, BP
  166, 38042 Grenoble Cedex 9, France.}

\begin{abstract}
  The recent observation of a insulator to metal transition
  (IMT)\cite{Loa01} in pure LaMnO$_3$ at 32 GPa and room
  temperature, well above the Neel temperature (145 K) and below the
  Jahn-Teller transition temperature (780 K), opens the way to a study
  of the role of the orbital degrees of freedom on the electronic
  structure in a stoichiometric material.

  In this paper we focus our attention in the orbital aspects of the
  insulator to metal transition. We use a model Hamiltonian for the
  $e_g$ orbitals of Mn that includes the on site Coulomb repulsion
  $U$, the hopping $t$, and its dependence with pressure. In order to
  include in an appropriate way the strong correlations induced by the
  dominant electron-electron interactions we introduce auxiliary
  fields (Slave Bosons,SB) to the description of the low energy
  states. We use the O-Mn distance ($d$) dependence of $t$ according
  to \cite{Harrison} and the pressure-$d$ relation from the
  experimental data to describe the evolution of the electronic
  structure with pressure.

  Our results confirm and make transparent the conclusion reached in
  previous ab-initio calculations: the inclusion of the Coulomb energy
  is necessary and constitutes an important factor enhancing the
  orbital polarization in these compounds.
\end{abstract}

\maketitle

\section{Introduction}
\label{sec:intro}

The discovery of colossal magneto-resistance in manganese perovskytes,
of relevance to spintronics and other technological applications, has
opened a new field to the study of highly correlated systems (HCS). In
the three dimensional compounds like La$_{1-x}$Sr$_x$MnO$_3$ or in the
two dimensional ones as La$_{2-2x}$Sr$_{2+2x}$Mn$_2$O$_7$ the
interplay of several degrees of freedom, charge, spin, orbit and
lattice displacements determine the physical properties of the
system. The interactions between the different degrees of freedom
cannot be reduced to perturbation theory as in other materials and the
disorder produced by alloying to obtain the metallic state does not
contribute to the clarification of theory nor to the interpretation of
experiments.

The recent observation of a insulator to metal transition
(IMT)\cite{Loa01} in pure LaMnO$_3$ at 32 GPa and room temperature,
well above the Neel temperature (145 K) and below the Jahn-Teller
transition temperature (780K), opens the way to a study of the role of
the orbital degrees of freedom on the electronic structure in a
stoechiometric material.

Two theoretical studies relevant to this matter have appeared in the
literature almost simultaneously after the publication of the
experimental result: both papers are based on local density
approximation+Hubbard U (LDA+U) approximation.

In the first one by Wei-go Yin et al.\cite{Yin06} the electronic
structure of the ground state of LaMnO$_3$ (antiferromagnetic A phase)
is analysed to determine the relative importance of the
electron-electron (e-e) against electron-lattice (e-l) Jahn-Teller
interactions. It concludes that the e-l interaction by itself is not
sufficient to stabilize the orbital ordered state and emphasizes the
importance of the e-e interaction to facilitate the Jahn-Teller
distortion.

In the second paper by Yamasaki et al.\cite{Yamasaki06} LDA+U and LDA+
dynamical mean field theories are used to analyze the metal insulator
transition taking place at 32 GPa in the paramagnetic phase of the
same material. The authors conclude that the transition at 32 GPa is
caused by the orbital splitting of the $e_g$ bands and that both e-e and
e-l interactions are needed to explain the insulating character of the
substance at lower pressures.

In this paper we focus our attention on the orbital aspects of the
insulator to metal transition. In order to include in an appropriate
way the strong correlations induced by the dominant e-e interactions
we follow Feiner and Ole\'s in introducing auxiliary fields (Slave
Bosons,SB) to the description of the low energy states.\cite{Feiner05}

We start by defining a simplified Hamiltonian, two parameters: $U$ on
site Coulomb repulsion, and $\Delta\varepsilon$ a measure of the Jahn
Teller splitting, to describe the $e_g$ states of the system and use
SB to calculate the lowest energy varying the parameters to obtain a
phase diagram. This procedure allow to identify the values of the
parameters that are appropriate to describe the transition, as well as
the thermodynamics and transport properties of LaMnO3 in the
paramagnetic phase.

\section{Methods}
\label{sec:methods}

\subsection{Hamiltonian}
\label{sec:H}

To describe the active electrons in LaMnO$_3$ we use the
double-exchange model that contains the essential physics of
manganites\cite{Popescu06}. The four 3$d$ electrons in each Mn$^{3+}$
site are polarized in the same direction due to the large Hund
coupling. Three of them occupy the $t_{2g}$ orbitals and are
considered localized forming a spin $S=3/2$, while the fourth
occupying the $e_g$ state is itinerant. We treat the localized spin
classically and since the fourth has to be parallel to the local spin
we can consider the $e_g$ electron as spinless with Hamiltonian
\begin{equation}
  \label{eq:H}
  H = H_t+H_U+H_{JT}
\end{equation}
where the first term, $H_t$, is the kinetic energy given by
\begin{equation}
  \label{eq:Ht}
  H_t = \sum_{\left \langle ij \right \rangle \alpha \beta } t_{ij}^{\alpha\beta} 
  c_{i\alpha}^{\dagger} c_{j\beta} 
\end{equation}
with $\alpha,\beta = 1,2$ corresponding to the (possibly site
dependent) orthogonal basis for the two $e_g$ orbitals. The values of
the hopping integrals $t_{ij}^{\alpha\beta}$ depend both on the type
of orbitals involved and on the direction between sites $i,j$. In
LaMnO$_3$, there is a staggered order in the $x$-$y$ plane and the
orbitals are stacked ferromagnetically along the $\hat{z}$ axis. The
dominantly occupied orbitals alternating in the $x$-$y$ plane are
$\left|x\right\rangle = \left|3x^2-r^2\right\rangle$ and
$\left|y\right\rangle = \left|3y^2-r^2\right\rangle$, which define
then the low energy orbitals $\left|1\right\rangle$ in each of the
sublattices, respectively $A$ and $B$. The corresponding higher energy
orbitals $\left| 2\right\rangle $ are therefore respectively $\left|
  y^{2}-z^{2}\right\rangle $ and $\left| x^{2}-z^{2}\right\rangle
$. The hopping parameters between these orbitals are the followings:

\begin{eqnarray}
  \label{eq:hoppings}
  \hat{t}_x^{AB} &=& t \left ( 
    \begin{array}{cc}
      1/2 & 0  \\
      -\sqrt{3}/2 & 0 
    \end{array} 
    \right ) \nonumber \\
  \hat{t}_y^{AB} &=& t \left ( 
    \begin{array}{cc}
      1/2 &  -\sqrt{3}/2 \\
      0 & 0 
    \end{array}  
    \right ) \nonumber \\
  \hat{t}_z^{AA} &=& \hat{t}_z^{BB} =t \left ( 
    \begin{array}{cc}
      -1/4 &  -\sqrt{3}/4 \\
      -\sqrt{3}/4 & -3/4 
    \end{array}  
    \right ) 
\end{eqnarray}
$t$ being the hopping between $\left| x\right\rangle $ ($\left|
  y\right\rangle $) orbitals along the $x$ ($y$) direction. 

The magnetic order is introduced by modulating the hopping integrals
by the factor $\exp(i A_{ij})\cos(\theta_{ij}/2)$,\cite{Dagotto} with
$\theta_{ij}$ being the angle between the $t_{2g}$ localized spins in
the neighboring sites $i,j$ and $A_{ij}$ a hopping phase. As we are
interested in the paramagnetic phase that appears at room temperature
$T\gg T_{N}$, we assume that the localized spins are completely random
and consider a mean field approximation in which this factor is
averaged giving a value $\langle \exp(i A_{ij})\cos(\theta_{ij}/2)
\rangle =2/3$. The $t_{ij}^{\alpha \beta }$ are the same than for a
ferromagnetic phase with a factor $2/3$. In the following we take this
renormalized hopping as the reference $t$.

The on-site Coulomb interaction between $e_g$ electrons occuping both
orbitals on the same site is given by
\begin{equation}
  \label{eq:HU}
  H_U = U \sum_{i} n_{i1} n_{i2}
\end{equation}
with $n_{i\alpha} = c_{i\alpha}^{\dagger} c_{i\alpha}$ the number
operators.

Finally, to model the effect of the Jahn-Teller (JT) deformation, we
add a term that shifts the on-site energies of the $e_g$ orbitals
$1,2$ in opposite directions
\begin{equation}
  \label{eq:HJT}
  H_{JT} = \Delta\varepsilon \sum_i (n_{i2} - n_{i1})
\end{equation}
which corresponds to a JT splitting of $2\Delta\varepsilon$.

\subsection{Slave bosons method}
\label{sec:sb}

In order to treat the Hamiltonian Eq.~(\ref{eq:H}), we used the slave
boson theory of Kotliar and Ruckenstein\cite{Kotliar86} adapted to our
case of two orbitals instead of the two projection of spin. Therefore,
we introduce new boson ($e_i, d_i, b_{i\alpha}$) and pseudofermion
($f_{i\alpha}$) operators. The boson numbers $e_{i}^{\dagger }e_{i}$,
$b_{i\alpha }^{\dagger }b_{i\alpha }$ and $d_{i}^{\dagger }d_{i}$
represent the projectors onto the possible states $\left|
  0_{i}\right\rangle \left\langle 0_{i}\right| $, $\left| \alpha
  _{i}\right\rangle \left\langle \alpha _{i}\right| $ and $\left|
  d_{i}\right\rangle \left\langle d_{i}\right| $ so that

\begin{equation}
e_{i}^{\dagger }e_{i}+b_{i1}^{\dagger }b_{i1}+b_{i2}^{\dagger
}b_{i2}+d_{i}^{\dagger }d_{i}=1
\end{equation}
and
\[
b_{i\alpha }^{\dagger }b_{i\alpha }+d_{i}^{\dagger }d_{i}=c_{i\alpha
}^{\dagger }c_{i\alpha }
\]

The original fermion operators are replaced by
\begin{eqnarray}
  \label{eq:c_sb}
  c^{\dagger}_{i1} &=&
  \left(b^{\dagger}_{i1}e_i+d^{\dagger}_ib_{i2}\right)f^{\dagger}_{i1} \nonumber\\
  c^{\dagger}_{i2} &=&
  \left(b^{\dagger}_{i2}e_i+d^{\dagger}_ib_{i1}\right)f^{\dagger}_{i2}
\end{eqnarray}
which correspond to a representation of the empty
($\left|0_i\right\rangle$), single occupied ($\left|1_i\right\rangle$,
$\left|2_i\right\rangle$) and doubled occupied ($\left|d_i\right\rangle$) local
states by
\begin{eqnarray}
  \label{eq:sb}
  \left|0_i\right\rangle &=& e^{\dagger}_i\left|vac\right> \nonumber\\
  \left|1_i\right\rangle &=& b^{\dagger}_{i1}f^{\dagger}_{i1}\left|vac\right> \nonumber\\
  \left|2_i\right\rangle &=& b^{\dagger}_{i2}f^{\dagger}_{i2}\left|vac\right> \nonumber\\
  \left|d_i\right\rangle &=& d^{\dagger}_{i}f^{\dagger}_{i2}f^{\dagger}_{i1}\left|vac\right>
\end{eqnarray}
with $\left|vac\right>$ correspondig to the vacuum state.

The anticommutation rules for the original fermions are guaranteed
provided the following constraints are satisfied 

% As in the original
% formulation, the enlarged Fock space contains unphysical states which
% must be eliminated by imposing the following constraints

\begin{equation}
  \label{eq:const}
  b^{\dagger}_{i\alpha}b_{i\alpha}+d^{\dagger}_id_i = 
  f^{\dagger}_{i\alpha}f_{i\alpha}{\rm \ \ \ \ \ \ \ }\alpha =1,2
\end{equation}
% \begin{eqnarray}
%   \label{eq:const}
%   e^{\dagger}_{i}e_{i}+b^{\dagger}_{i1}b_{i1}+
%   b^{\dagger}_{i2}b_{i2}+d^{\dagger}_id_i = 1 \nonumber\\
%   b^{\dagger}_{i1}b_{i1}+d^{\dagger}_id_i = f^{\dagger}_{i1}f_{i1} \nonumber\\
%   b^{\dagger}_{i2}b_{i2}+d^{\dagger}_id_i = f^{\dagger}_{i2}f_{i2}
% \end{eqnarray}
which are implemented by means of the corresponding Lagrange
multipliers $\left\{\lambda_i, \mu_{i1}, \mu_{i2}\right\}$. To recover
the correct result in the uncorrelated ($U=0$) limit, a
renormalization of the bosonic factor in Eq.~(\ref{eq:c_sb}) is
necessary. In analogy with the spin case, the renormalized bosons
factors take the form
\begin{eqnarray}
  \label{eq:z}
  z^{\dagger}_{i1} &=& \frac{b^{\dagger}_{i1}e_i+d^{\dagger}_ib_{i2}}
  {\sqrt{\left(1-e^{\dagger}_ie_i-b^{\dagger}_{i2}b_{i2}\right)
      \left(1-d^{\dagger}_id_i-b^{\dagger}_{i1}b_{i1}\right)}} \nonumber\\
  z^{\dagger}_{i2} &=& \frac{b^{\dagger}_{i2}e_i+d^{\dagger}_ib_{i1}}
  {\sqrt{\left(1-e^{\dagger}_ie_i-b^{\dagger}_{i1}b_{i1}\right)
      \left(1-d^{\dagger}_id_i-b^{\dagger}_{i2}b_{i2}\right)}}
\end{eqnarray}

With this slave boson representation, the Hamiltonian (\ref{eq:H})
reduces to
\begin{eqnarray}
  \label{eq:H_sb}
  H &=& \sum_{\left < ij \right > \alpha \beta } t^{\alpha\beta}_{ij} 
  z^{\dagger}_{i\alpha} z_{j\beta} f^{\dagger}_{i\alpha} f_{j\beta}
  \nonumber\\
  && + U \sum_i d^{\dagger}_id_i 
  + \Delta\varepsilon \sum_i (n_{i2} - n_{i1})
  \nonumber\\
  && + \sum_{i\alpha} \mu_{i\alpha}
  \left(f^{\dagger}_{i\alpha}f_{i\alpha}-
    b^{\dagger}_{i\alpha}b_{i\alpha}-d^{\dagger}_id_i\right)
  \nonumber\\
  && + \sum_{i} \lambda_i \left(e^{\dagger}_{i}e_{i}+
    b^{\dagger}_{i1}b_{i1}+b^{\dagger}_{i2}b_{i2}+
    d^{\dagger}_id_i-1\right)
\end{eqnarray}
where the pseudo-fermions number operators are $n_{i\alpha }=f_{i\alpha
}^{\dagger }f_{i\alpha }=c_{i\alpha }^{\dagger }c_{i\alpha }$. We have
studied the solutions of this Hamiltonian in the mean-field approximation,
in which we replace the boson operators by their averages obtained from the
minimization of the band energy. In pure LaMnO$_{3}$ the number of
conduction electrons is $n=1$ and in the absence of charge ordering $%
n_{i1}+n_{i2}=n_{i}=1$.\ In this case the band renormalization factor is
independent of the type of orbitals involved $q=z_{i\alpha }^{\dagger
}z_{j\beta }$. 

\section{Results}
\label{sec:results}

\begin{figure}[ht]
  \begin{center}                          
    \includegraphics*[width=6.5cm]{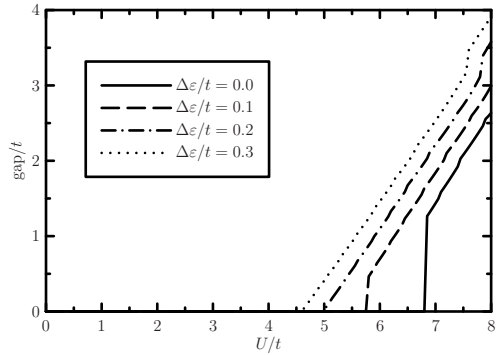}
  \end{center}                                                 
\caption{Dependence of the band gap as a function of $U$.}
\label{fig:gap_U}                                            
\end{figure}

To characterize the solutions of Eq.~(\ref{eq:H_sb}) for different
values of the two parameters $U$ and $\Delta\varepsilon$, we
considered both the magnitude of the pseudofermion band gap and the
orbital polarization. In Fig.~\ref{fig:gap_U} we show the dependence
of the band gap with the value of $U$. We can see that for
$\Delta\varepsilon=0$ there is a jump in the band gap at a critical
value $U_c$, corresponding to a first order metal-insulating
transition (in the parameter $U$). When we begin to increase the value
of the JT splitting, $U_c$ shifts to lower values and also the initial
band gap in the insulating phase also diminishes. Beyond certain value
of $\Delta\varepsilon$ (between $0.225t$ and $0.250t$), the
metal-insulating transition is no more of first order, and the band
gap opens smoothly.

\begin{figure}[ht]
  \begin{center}                          
    \includegraphics*[width=6.5cm]{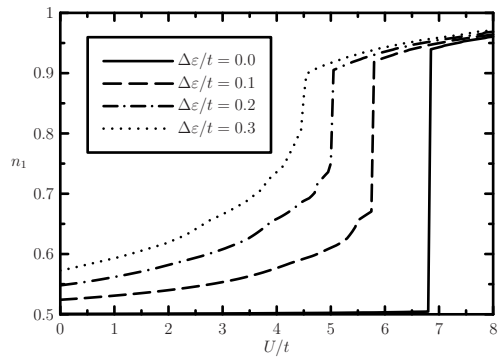}
  \end{center}                                                 
  \caption{Dependence of the orbital polarization as a function of
    $U$.}
\label{fig:n1_U}                                            
\end{figure}

Similar features can be observed in the orbital polarization shown in
Fig.~\ref{fig:n1_U}, where for small values of $\Delta\varepsilon$
there is a jump in the polarization at the same critical value $U_c$.
From these results, we can see that an insulating phase with orbital
polarization is possible even without a JT deformation. However, an
orbital polarization will induce, due to the electron-lattice
interaction, a JT deformation. We note also that for
$\Delta\varepsilon=0$ the orbital polarization is symmetric, i.e.  the
low energy orbital can be any combination of $e_g$ orbitals, and the
actual one will depend on the electron-lattice interaction.

\begin{figure}[ht]
  \begin{center}                          
    \includegraphics*[width=7cm]{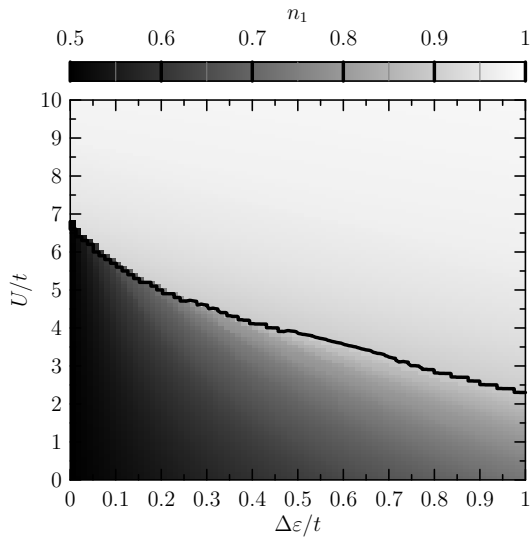}
  \end{center}                                                 
  \caption{Phase diagram $U$-$\Delta\varepsilon$. Solid line:
    metal-insulating transition. Gray tone: occupancy of the low
    energy orbital $n_1$.}
\label{fig:diag_Ude}
\end{figure}

In Fig.~\ref{fig:diag_Ude} we show the phase diagram
$U-\Delta\varepsilon$ where the line corresponds to the value of $U_c$
at the metal-insulating transition, and the gray tone represent the
orbital polarization given by the occupancy of the low energy orbital
$n_1$. While in the insulating phase there is always a high orbital
polarization, in the metallic phase the amount of orbital polarization
depends on $\Delta\varepsilon$: the higher the value of
$\Delta\varepsilon$, the higher the polarization. The critical value
is close to $U_c = 7t$ for the case without JT splitting, and descends
to almost $2t$ for a splitting $\Delta\varepsilon = t$.

\begin{figure}[ht]
  \begin{center}                          
    \includegraphics*[width=6cm]{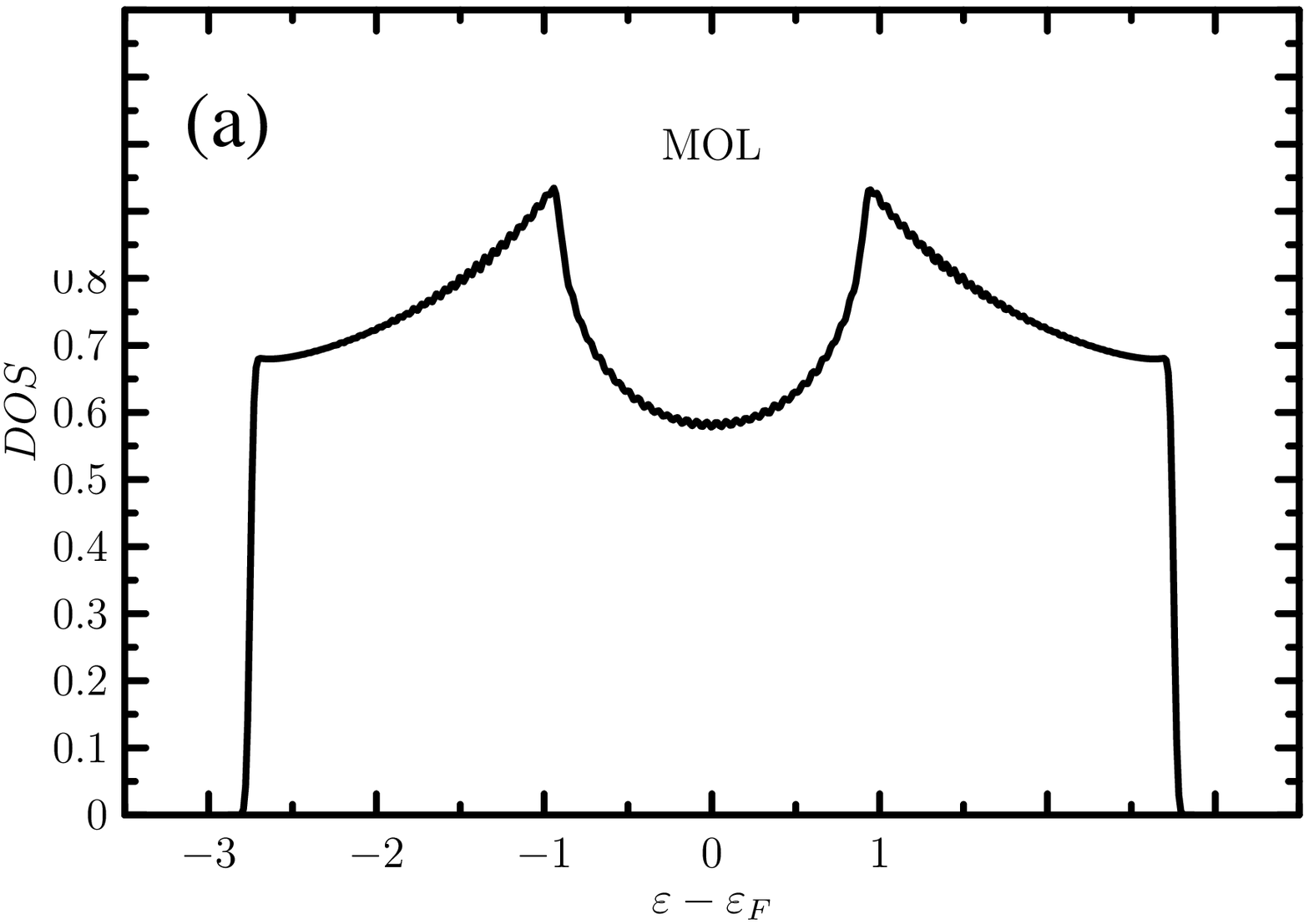}
    \includegraphics*[width=6cm]{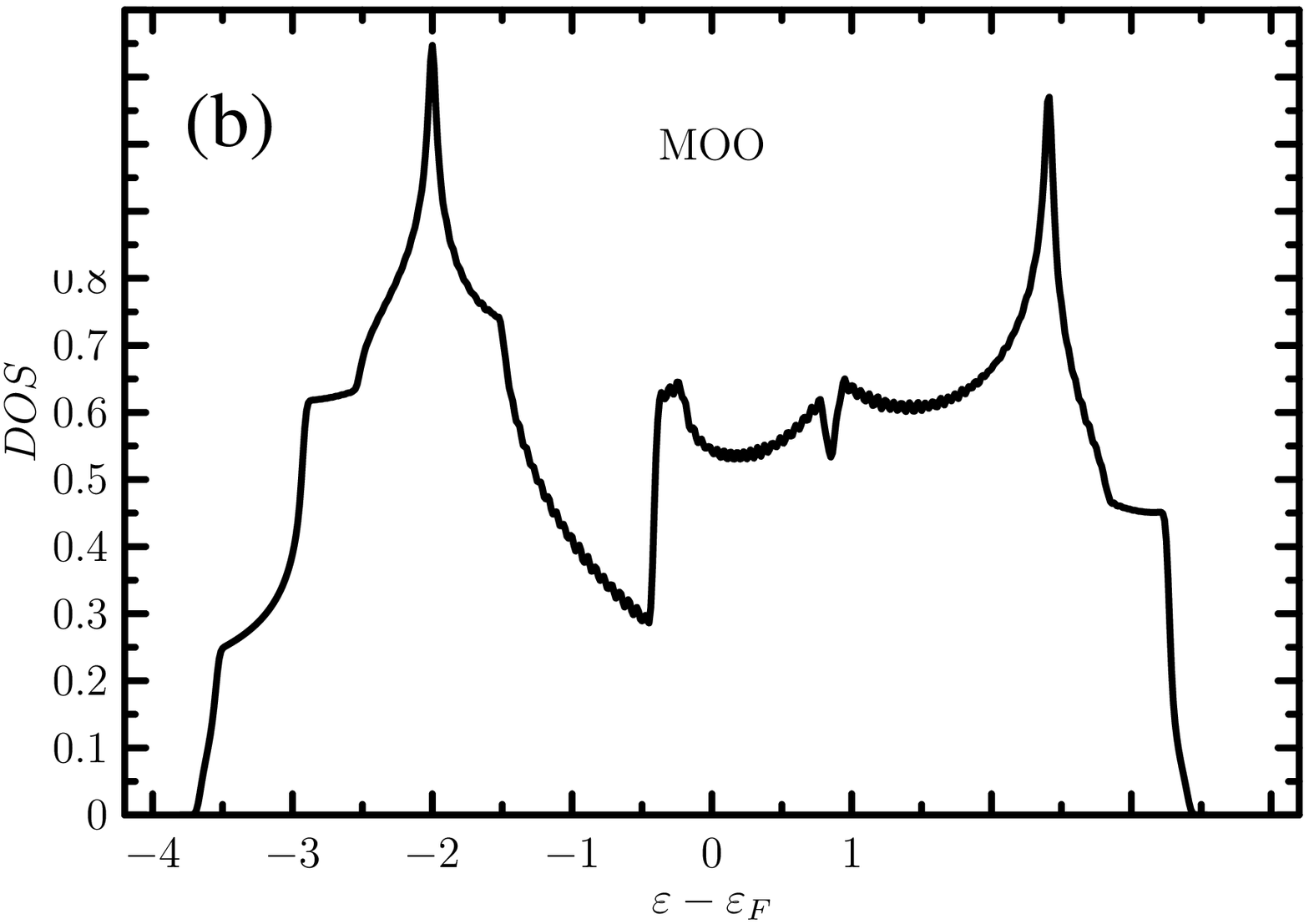}
    \includegraphics*[width=6cm]{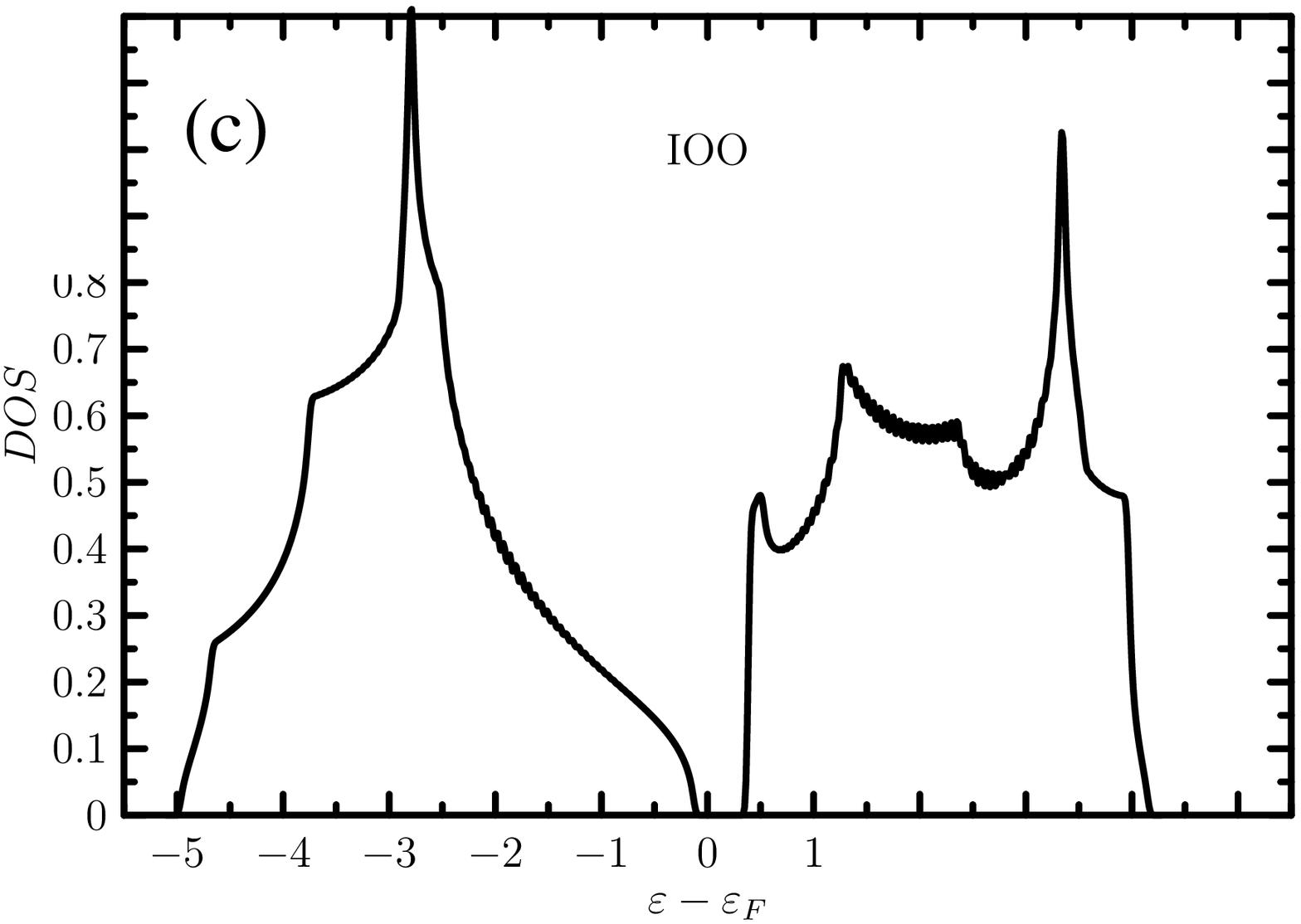}
  \end{center}                                                 
  \caption{DOS for the different phases: (a) metallic phase with very
    small orbital polarization, (b) metallic phase with orbital
    polarization, and (c) insulating phase.}
\label{fig:DOSs}
\end{figure}

To better characterize the different phases we have calculated the
pseudofermion density of states (DOS). In Fig.~\ref{fig:DOSs} we show
this DOS for three different representative parameters in the phase
diagram: 
\begin{itemize}
\item ($U=3.5t$, $\Delta\varepsilon=0.0 t$): metallic phase with very
  small orbital polarization which we have named orbital liquid (MOL).
  In this case, there is almost no orbital polarization ($n_1 =
  0.502$) and the renormalization factor is $q=0.917$.
\item ($U=3.5t$, $\Delta\varepsilon=0.4 t$): metallic phase with
  orbital order (MOO). The orbital polarization is now higher ($n_1 =
  0.764$) and the renormalization factor is closer to one ($q=0.941$).
\item ($U=3.5t$,$\Delta\varepsilon=0.8 t$): insulating phase, also
  with orbital order (IOO). In this case there system is almost fully
  polarizad with $n_1=0.920$ and also as a consequence $q=0.986$ is
  also close to one.
\end{itemize}

To simulate the effect of pressure on the electronic properties, we
take the experimental data on the IM transition and model the
dependence of the parameters with the pressure. All magnitudes are
given in terms of the effective hopping at $P=0$, which is of the
order of $t_0 \simeq 0.4$ eV\cite{Yin06}. The JT splitting is taken to
vary linearly from a value of $\Delta\varepsilon = 1.85
t_0$\cite{Yamasaki06} at $P=0$ to $\Delta\varepsilon = 0$ at the
experimental JT suppression pressure $P=18$ GPa. The {\it e-e}
interaction constant is considered constant. Finally, the effective
hopping has a dependence of the form\cite{Harrison}
\begin{equation}
  \label{eq:t_P}
  t(P) = t_0 \left({d_0 \over d(P)}\right)^7
\end{equation}
where $d(P)$ is the mean Mn-O distance as a function of the pressure,
taken from the experimental data, and $d_0 = d(0)$.

\begin{figure}[ht]
  \begin{center}                          
    \includegraphics*[width=7cm]{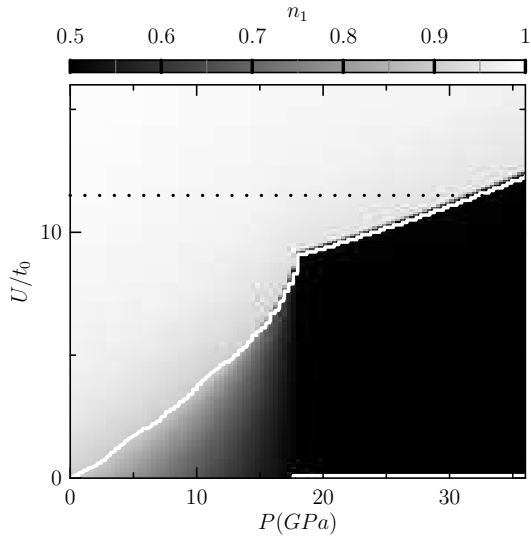}
  \end{center}                                                 
  \caption{Phase diagram $U$-$P$. Solid line: metal-insulating
    transition. Dot line: $U=11.5 t_0$. Gray tone: occupancy of
    the low energy orbital $n_1$. }
\label{fig:diag_UP}
\end{figure}

From this dependence of the parameters with the presure, in
Fig.~\ref{fig:diag_UP} we show the phase diagram $U-P$ where as before
the solid line corresponds to the value of $U_c$ at the
metal-insulating transition, gray tone represent the orbital
polarization given by the occupancy of the low energy orbital
$n_1$. Note that the values of $U$ in the vertical axis are in units
of $t_0$, and we have added a dot line showing that for the estimated
value of $U=11.5 t_0$ we obtain a insulator to metal transition pressure
close to the experimental one of $P\simeq 32$ GPa.

\section{Conclusions}
\label{sec:conc}

We use a minimum parameters model Hamiltonian to study the evolution
of orbital polarization in LaMnO$_3$. In order to include
appropriately the effects of correlation, we resort to the Slave
Bosons technique which was previously used by Feiner and Oles in the
context of manganites.

We calculate the electronic structure and from it, the electronic
energy to obtain a phase diagram in terms of two independent
parameters $U/t$ an $\Delta\varepsilon/t$. The results can be
translated to the effect of pressure on the material by modelling the
variation of the parameters with volume and connecting to pressure
though the compressibility.

The same Hamiltonian could be used to represent the orbital state of
other compounds where trivalent Rare Earths substitute partially or
totally La, again though modelling of the variation of the hopping ,
$U$ or $\Delta\varepsilon$ as an effect of substitution and pressure,
as for example in equation \ref{eq:t_P}.

Our results confirm and make transparent the conclusion reached in
previous ab-initio calculations: The inclusion of the Coulomb energy
is necessary and constitute an important factor enhancing the orbital
polarization in these compounds.
 
From the density of states it is possible to calculate the gap in the
insulating phases and the number of carriers as a function of
temperature and pressure, in order to compare with the results of Loa
et al.\cite{Loa01}

A natural continuation of these results is the evaluation of the
elastic energy involved in the Jahn-Teller distortion that will appear
at the polarized phases, the results that will be published elsewhere.

We acknowledge support by CONICET.

\end{document}